\documentclass[preprint,prb,amsmath,superscriptaddress,aps,subeqn]{revtex4}
\usepackage{graphicx}

\begin{document}

\title{First principles calculations of lattice thermal conductivity in mono- and bi-layer graphene}

\author{B. D. Kong}
\affiliation{Department of Electrical and Computer Engineering,
North Carolina State University, Raleigh, NC 27695-7911}

\author{S. Paul}
\affiliation{Department of Physics, North Carolina State
University, Raleigh, NC 27695-8202}

\author{M. Buongiorno Nardelli} \email{mbnardelli@ncsu.edu}
\affiliation{Department of Physics, North Carolina State
University, Raleigh, NC 27695-8202}
\affiliation{Computer Science and Mathematics Division, Oak Ridge National Laboratory, Oak Ridge, TN 37831}

\author{K. W. Kim} \email{kwk@ncsu.edu}
\affiliation{Department of Electrical and Computer Engineering,
North Carolina State University, Raleigh, NC 27695-7911}

\begin{abstract}
Using calculations from first principles we have investigated the
lattice thermal conductivity of ideal mono- and bi-layer graphene
sheets. Our results demonstrate that the intrinsic thermal conductivity of both mono- and bi-layer
graphene is around 2200 Wm$^{-1}$K$^{-1}$ at 300 K, a value close to the one observed theoretically and
experimentally in graphite basal plane,
and at higher temperatures it decreases with the expected $T^{-1}$
dependence. The little variation between mono- and bi-layer thermal
conductivity suggests that increasing the number of layers does not
affect significantly the in-plane thermal properties of these
systems.
\end{abstract}

\maketitle

Although carbon-based low dimensional materials have been subject of
intense investigations for decades, the recent synthesis of
graphene, a single 2-dimensional sheet of hexagonal carbon in the
honeycomb lattice, has opened  new avenues of research in the search
of novel, alternative materials for
microelectronics.~\cite{RiseofGraphene} The peculiar electronic
structure of  graphene~\cite{RiseofGraphene} implies that electrons
have a negligible effective mass around the K point of the Brillouin
zone (Dirac point) and thus a high electrical conductivity. However,
electric currents are inherently coupled with Joule heating so it is
imperative to investigate the thermal property of these systems to
insure proper operations of an eventual device. Usually a good
electric conductor is a good thermal conductor. However, this
expectation is broken in the case of diamond, another kind of
carbon-base crystal, which exhibit a high thermal conductivity
regardless of its large band gap and low electric conductivity.
Graphite, the closest relative of graphene in the family of
carbon-based materials, displays high thermal conductivity along the
basal plane and, not surprisingly,  two orders of magnitude smaller
thermal conductivity along the \emph{c}-crystallographic
axis.~\cite{Klemens_carbon}
\par There is a relatively large scattering in the experimental data for
the thermal conductivity of highly oriented graphite and graphene
systems. Ref.~\onlinecite{Klemens_carbon} reports values of thermal
conductivity in the range $ 1660 - 1880$ Wm$^{-1}$K$^{-1}$. Balandin
\emph{et al.}~\cite{Balandin_Ex} measured values around $4840 -
5300$ Wm$^{-1}$K$^{-1}$ for a suspended graphene mono-layer. More
recently, Nika \emph{et al.}~\cite{Balandin_PrePrint} reported their
theoretical investigation on the thermal conductivity of mono-layer
graphene and publish values in a range between $3000 - 6500$
Wm$^{-1}$K$^{-1}$, depending on the choice of Gr\"{u}neisen
parameters, $\gamma$, which had values ranging from 0.8 to 2.0 as
deduced from Ref~\onlinecite{Marzari1}. This particular study was
based on the use of a valence force field method which resolves all
interatomic forces into bond-stretching and bond bending
modes.~\cite{Balandin_PrePrint,VFF}  and the Gr\"{u}neisen
parameters were assumed to be
constant for each phonon branch.~\cite{Balandin_PrePrint} Comparable values of thermal
conductivity have also been observed in Carbon Nanotubes (CNT): Kim
\emph{et al.}~\cite{CNT_Ex1} reported approximately 3000
Wm$^{-1}$K$^{-1}$ as a thermal conductivity of a multi-wall CNT with
140 \AA ~ diameter. Pop \emph{et al.}~\cite{CNT_Ex2} observed $2800
- 3900$ Wm$^{-1}$K$^{-1}$ for  single-wall CNT with diameters between
$ 0.5 - 10 $ $\mu$m at room temperature. Both experiments were
performed on a single suspended nanotube by heating it with
electrical current.
\par In order to resolve this apparent spread in
the values of thermal conductivity and understand the interplay between the number of
layers and the thermal response of these systems we have
investigated the intrinsic lattice thermal conductivity of ideal
mono- and bi- layer graphene using phonon dispersion relations and
lattice anharmonicity parameters calculated from first principles
methods. Our results show that mono- and bi- layer graphene sheets
have an intrinsic thermal conductivity superior to usual bulk solids
except diamond which has 600 $\sim$ 2000 Wm$^{-1}$K$^{-1}$
~\cite{diamond}.

\par The lattice thermal conductivity $\mathbf{\kappa}$ of a crystal
at finite temperature \emph{T} can be written as
\begin{equation}
\mathbf{\kappa}=\sum_{\lambda}
\mathbf{\kappa}_{\lambda}\label{eq:thcond1}
\end{equation}
\begin{equation}
\mathbf{\kappa}_{\lambda}=\sum_{\mathbf{q}}(\mathbf{v(q)}\cdot\mathbf{t})^2
\tau(\mathbf{q})C_{ph}(\omega)\label{eq:thcond2}
\end{equation}
where $\mathbf{t}$ is a unit vector in the direction of the thermal
gradient $\nabla\emph{T}$, $\mathbf{v(q)}$ is the group velocity of
the phonon modes, and $\tau(\mathbf{q})$ is their life
time.~\cite{Klemens_thermal,Ziman,Kim} Here, the index $\lambda$
runs over the phonon modes and $C_{ph}(\omega)$ is the contribution
of phonon modes to the specific heat whose form is
\begin{equation}
C_{ph}(\omega)=\hbar \omega \frac{dN^0}{d
\emph{T}}=\frac{(\hbar\omega)^2 }{k_B \emph{T}^2}
\frac{\exp(\hbar\omega/k_B \emph{T})}{[\exp(\hbar \omega/k_B
\emph{T})-1]^2}\label{eq:thcond3}
\end{equation}
where $k_B$ is Boltzmann constant and $\hbar$ is Plank constant. In
order to evaluate accurately the lattice thermal conductivity one
needs to obtain the phonon group velocity $\mathbf{v(q)}$ from
realistic phonon dispersion relations and to estimate the phonon
life time from a careful consideration of the possible relaxation
mechanisms.

\par As shown very elegantly in Ref~\onlinecite{Klemens_carbon}, one
can calculate the graphite thermal conductivity along the basal
plane considering only the LA and TA branches. The assumption is
that the phonon dispersion of graphite has cylindrical shape and the
ZA modes strongly interact only along the \emph{c}-direction, mainly
due to the large spacing and the weak bonding between the layers.
Since the group velocities of the optical modes are considerably
smaller than those of the acoustic modes, optical branches can be
disregarded. Therefore, we can consider only TA and LA modes in the
range above the temperature which corresponds to the highest ZA
frequency and we can replace the first Brillouin zone (FBZ) with a
circular cylinder and define the average sound velocity $\langle
v\rangle$ for the two-dimensional phonon gas consisting of LA and TA
modes as: $2/\langle v\rangle^2 = 1/\langle v_{LA}\rangle^2
+1/\langle v_{TA}\rangle^2$. Recent first principles calculations of
phonon dispersion in graphite~\cite{Marzari1,Marzari2} and our own
results validate this approach since the phonon dispersion curves
along the $\Gamma-A$ axis are almost flat, implying that the FBZ of
graphite looks like a circular column. Graphene sheets inherently
have a two dimensional nature so it seems reasonable to accept these
assumptions for the high temperature range in which the role of ZA
modes is not significant. According to our calculations, the maximum
of ZA mode is at 535 cm$^{-1}$ . This frequency corresponds to 123 K
($k_B T= \hbar \omega$), validating our assumption for any T larger
than that.

\par We evaluated the average phonon life time as limited by the
anharmonicity of lattice vibrations since this is the most
fundamental limiting factor which is not relying on the purity of
crystal or boundary termination technology. The analytical
expression for the relaxation rate from the anharmonic three-phonon
processes are readily derived as:
\begin{equation}\label{tau}
\frac{1}{\tau_{\lambda}}=2\gamma_{\lambda}^2 \frac{k_B T}{M v^2}
\frac{\omega^2}{\omega_m}
\end{equation}
where $\gamma_{\lambda}$ are Gr\"{u}neissen parameters, $k_B$ is
Boltzmann constant, $T$ is temperature, $M$ is the mass of atoms and
$\lambda$ runs over the phonon
modes.~\cite{Klemens_carbon,Klemens_thermal,Klemens_solid} $\omega_m$
and $v$ are Debye frequency and the average sound velocity.

\par The phonon dispersion relations used in this study have been
obtained from first principles calculations within Density
Functional Theory and Density Functional Perturbation Theory. The
predictive power of this approach has been demonstrated by
the excellent agreement between theoretical predictions and
experimental observations in recent studies on the lattice
vibrational modes of graphite and graphene.~\cite{Marzari1,Marzari2}
In this study, the calculation for phonon dispersion and
Gr\"{u}neisen parameters for the mode anharmonicity were obtained by
using the \textsc{PWscf} package of the \textsc{QUANTUM-ESPRESSO}
distribution.~\cite{QuanEspr}

\par Fig.~\ref{fig:dispersion} shows the calculated phonon
dispersions along the high symmetry lines. The phonon frequency
values of high symmetry points such as $\Gamma$, M, and K can be
compared with previous studies and the results for mono-layer
graphene are summarized in Table~\ref{tab:table1}. The  results for
bi-layer graphene  show a very similar behavior. As shown in the
figure, another optical vibrational mode appears at $\Gamma$ of
bi-layer graphene phonon dispersion.  This ZO' mode at 78 cm$^{-1}$
is produced by the displacement of the carbon atoms in opposite
directions along the {\it c} crystalline axis (optical mode). In
graphite the same mode has a frequency of 95
cm$^{-1}$.~\cite{GraphiteExp1} The agreement of our data with the
previous theoretical and experimental studies is clearly excellent.
From these accurate phonon dispersion curves we calculated the group
velocity $\mathbf{v(q)}$ and the Gr\"{u}neissen parameter
$\gamma_{\lambda}$ for each phonon mode $\lambda$. The latter are
defined as the negative logarithmic derivative of the frequency of
the mode with respect to volume: $\gamma_{\lambda} =
-[a/2\omega_{\lambda}
(\mathbf{q})][d\omega_{\lambda}(\mathbf{q})/da]$~\cite{Ashcroft,Marzari1}
where $a$ is the lattice constant. By calculating phonon dispersion
relation with small deviation from the original lattice constant and
using the above definition, we computed the Gr\"{u}neissen
parameters over the FBZ as shown in Fig.~\ref{fig:gruneisen}. In
mono-layer graphene,  the ZO and ZA modes have negative
$\gamma_{\lambda}$, between  -1.38 and -0.17 for ZO, and between -53
and -1.46, for ZA. For the other modes, the values show variations
between 0.16 and 2.76. Positive values of $\gamma$ correspond to a
decrease of the frequency of the modes as the lattice parameter
increases: the atoms are less tightly bound to their equilibrium
position. Inversely, negative values of $\gamma$ correspond to an
increase of the frequencies upon the expansion of the lattice. Since
ZA and ZO are out-of-plane transverse polarization modes, one can
imagine that if a graphene sheet is expanded, the force to move the
atoms in the vertical direction will increase.  As for the bi-layer,
the dispersion relations for the Gr\"{u}neissen parameters display
an additional contribution corresponding to the ZO' mode. In
addition, $\gamma_{\rm ZA}$ and $\gamma_{\rm ZO'}$ of of the
bi-layer show slightly different behavior at the $\Gamma$ point:
they are positive. Small values of $\mathbf{q}$ (around $\Gamma$)
correspond to long wavelength phonons and a positive value of
$\gamma$ means that the atoms are less tightly bound. The variation
of $\gamma$ from negative to positive tells us that the properties
of the ZA or ZO' modes' collective movement are different for long
wavelength and short wave length: as the volume is increased, and
with it the intra-layer distance, $\omega_{ZA,ZO'}$  decreases at
$\Gamma$. Atoms in both layers loose some of  their coherence in the
long wavelength vibrational modes. There are two TA and LA modes so
the mean value of $\gamma_{\rm TA1}$, $\gamma_{\rm TA2}$,
$\gamma_{\rm LA1}$, and $\gamma_{\rm LA2}$ in the FBZ was found to
be 0.52, 0.53, 1.58, and 1.56, respectively.

\par The normalized thermal conductivity $\kappa/\tau$ is reported in
Fig.~\ref{fig:kappatau}. The curves are obtained from the {\it ab
initio} phonon dispersion relations and the two-dimensional phonon
gas model for infinite mono- and bi-layer graphene sheets along
(100) direction from 200 K to 500 K. For mono- and bi- layer, we
observe a similar behavior over the calculated temperature range:
both show almost the same normalized thermal conductivities. For
instance, at $T$=300 K, $\kappa/\tau = 4.78 \times10^{13}
Wm^{-1}s^{-1}$ for mono-layer and $\kappa/\tau = 4.83 \times10^{13}
Wm^{-1}s^{-1}$. In both cases the contribution from the LA mode is
larger than the one from the TA mode due to the large group velocity
of the LA mode. From the data in Fig.~\ref{fig:kappatau}(b), it is
evident that the magnitude of the TA and LA modes is decreased when
another layer is added to the system; however, also the number of
modes doubles so that  the normalized conductivity is the same.
Moreover, going from mono- to bi-layer graphene, the slope of the TA
and LA branches do not change significantly. However, the thickness
of the layer doubles, so the contribution from each branch becomes
one half of the value of the mono-layer.~\cite{thickness1}

\par The lattice thermal conductivity with phonon mean life time
limited by lattice anharmonicity effects is presented in
Fig.~\ref{fig:kappa} for the same temperature range of
Fig~\ref{fig:kappatau} and along the (100) direction. Both mono- and
bi- layer graphene display a similar $\sim 1/T$ dependence at high
temperature. The values of $\kappa$ at room temperature (300 K)
differ only by 8 Wm$^{-1}$K$^{-1}$ . As previously mentioned, the
contribution from the LA branch is larger than that from the TA
branch in the case of normalized thermal conductivity. However, as
shown in the graphs, the TA mode transfers a larger portion of
thermal energy than the LA mode in the real thermal conductivity.
This is due to the longer phonon life time of TA which is expected
from the comparison of the Gr\"{u}neissen parameters. The LA mode
has three times larger $\gamma$. Overall, nine times longer phonon
mean life time is expected for TA so that this longer life time
compensate the smaller group velocity. Klemens \emph{et
al.}~\cite{Klemens_carbon} predicted 1900 Wm$^{-1}$K$^{-1}$ as the
lattice thermal conductivity along the basal plane for the bulk
graphite at 300 K. Our results substantially agree with this
previous estimate. The small difference, around 300
Wm$^{-1}$K$^{-1}$ at room temperature, that we find can be explained
by two reasons. First, in Ref. \onlinecite{Klemens_carbon} they
assumed $\gamma$=2 for both modes, a choice that eventually
decreases the thermal conductivity. Second, they assumed an average
sound velocity in the FBZ which could increase the thermal
conductivity since the actual group velocity of phonon modes
decreases when $\mathbf{q}$ approaches the zone boundary. The Debye
frequencies $\omega_m$ of two dimensional phonon gas in mono- and
bi- layer graphene are 1265 cm$^{-1}$ and 1243 cm$^{-1}$
respectively and those almost correspond to the maximum frequencies
of the LA mode at the zone boundary which supports the assumption
that the frequency below $\omega_m$ (TA and LA) contributes to the
thermal conduction. Finally, in Fig.~\ref{fig:kappa} we present the
angular dependence of the thermal conduction. $\theta$ in the figure
is the angle between the direction of the thermal gradient and the
(110) crystalline direction. The angular dependence was calculated
for different thermal gradient directions where, however, we observe only a
negligible variation (less than 0.1\% at 300 K).

\par When it comes to realistic finite graphene nanoribbon, other
possible scattering mechanisms can arise, in particular scattering
by impurities  and by the discontinuous structure of the ribbon side
terminations. To evaluate these effects one should both measure the
density of substitutional impurity atoms and the mass difference
between the impurity atoms and the carbon atoms and gain a more
complete understanding of the geometry of the system. Both pieces of
information at not yet readily available from the current status of
research on these systems. Therefore, our values for the intrinsic
lattice thermal conductivity of mono- and bi- layer graphene can be
regarded as a theoretical upper limit of the thermal conductivity of
more realistic graphene nanoribbons.

This work was supported, in part, by the NERC/NIST SWAN-NRI
research center and the DARPA/HRL CERA program.  MBN wishes
to acknowledge partial support from the Office of Basic Energy Sciences, U.S. Department of Energy at
Oak Ridge National Laboratory under contract DE-AC05-00OR22725 with UT-Battelle, LLC.
Calculations have been carried out at the Center for Computational Sciences at ORNL and
the NC State University HPC and Grid Computing initiative.

\clearpage \noindent Figure Captions


\vspace{0.5cm} \noindent Figure~1(a,b). (Color online) Phonon
dispersion of (a) a mono-layer and (b) a bi-layer graphene sheet
along the high symmetry line.

\vspace{0.5cm} \noindent Figure~2(a,b,c). (Color online)
Gr\"{u}neisen parameter of (a) a mono-layer and (b) a bi-layer
graphene sheets along the high symmetry lines. (c) Two dimensional
surface of Gr\"{u}neisen parameter of a mono-layer graphene sheet
over the entire FBZ.

\vspace{0.5cm} \noindent Figure~3.(a,b) (Color online)  Normalized
thermal conductivity in (100) direction of (a) a mono-layer and (b)
a bi-layer graphene sheet.

\vspace{0.5cm} \noindent Figure~4.(a,b) (Color online) Thermal
conductivity in (100) direction of (a) a mono-layer and (b) a
bi-layer graphene sheet. Inset of (a): the angular dependence of the
thermal conductivity of a mono-layer graphene sheet at T = 300 K.
$\theta$ is the angle between the thermal gradient and the (110)
direction.

\clearpage
\begin{table*}
\caption{\label{tab:table1}Comparison of the phonon frequencies of
graphene at $\Gamma$, M, and, K in cm$^{-1}$ from various
studies.}
\begin{ruledtabular}
\begin{tabular}{ccccccccccccccc}
&$\Gamma_{\rm ZO}$&$\Gamma_{\rm LO/TO}$&M$_{\rm ZA}$&M$_{\rm
TA}$&M$_{\rm ZO}$&M$_{\rm LA}$&M$_{\rm LO}$&M$_{\rm TO}$&K$_{\rm
ZA}$
&K$_{\rm ZO}$&K$_{\rm TA}$&K$_{\rm LA}$&K$_{\rm LO}$&K$_{\rm TO}$\\
\hline This
study~\cite{QEdetail}&884&1560&473&627&641&1318&1360&1396&532
&550&997&1210&1228&1327\\

Theoretical\footnotemark[1]&881&1554&471&626&635&1328&1340&1390&535
&535&997&1213&1213&1288\\
Experimental&861\footnotemark[2]&1590\footnotemark[2]&465\footnotemark[2]&630\footnotemark[2]&670\footnotemark[2]&1290\footnotemark[3]
&1321\footnotemark[3]&1389\footnotemark[4]&482\footnotemark[4]
&588\footnotemark[4]&&1184\footnotemark[4]&1184\footnotemark[4]&1313\footnotemark[3]\\
\end{tabular}
\end{ruledtabular}
\footnotetext[1]{Reference~\onlinecite{Marzari1}. Graphene sheets}
\footnotetext[2]{Reference~\onlinecite{GraphiteExp1}. Graphite}
\footnotetext[3]{Reference~\onlinecite{GraphiteExp2}. Graphene
sheets} \footnotetext[4]{Reference~\onlinecite{GraphiteExp3}.
Graphite}
\end{table*}

\clearpage
\begin{center}
\begin{figure}
\vspace*{0.5 in}
\includegraphics*[width=5.0in]{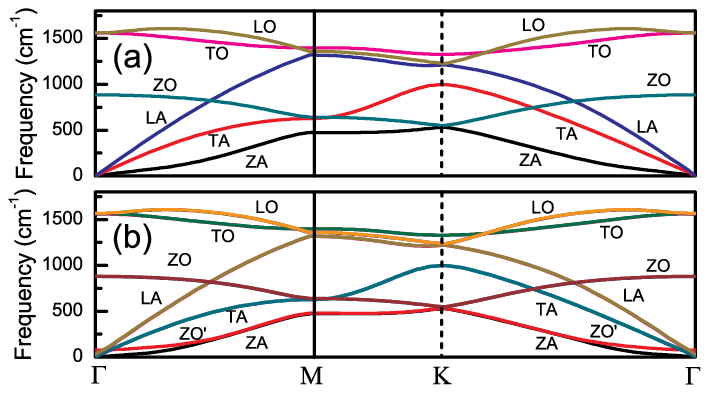}
\label{fig:dispersion}
\end{figure}
\vspace{1in} { Fig. 1(a,b): Kong \emph{et al.}}
\end{center}

\clearpage
\begin{center}
\begin{figure}
\vspace*{0.5 in}
\includegraphics*[width=5.0in]{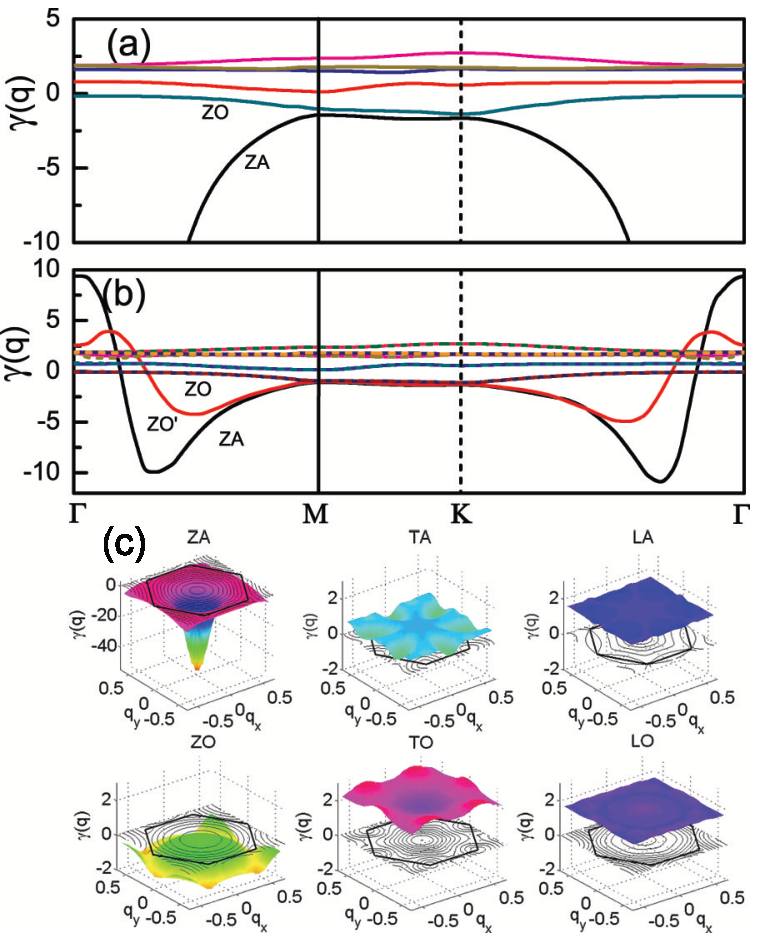}
\label{fig:gruneisen}
\end{figure}
\vspace{1in} { Fig. 2(a,b,c): Kong \emph{et al.}}
\end{center}

\clearpage
\begin{center}
\begin{figure}
\vspace*{0.5 in}
\includegraphics*[width=5.0in]{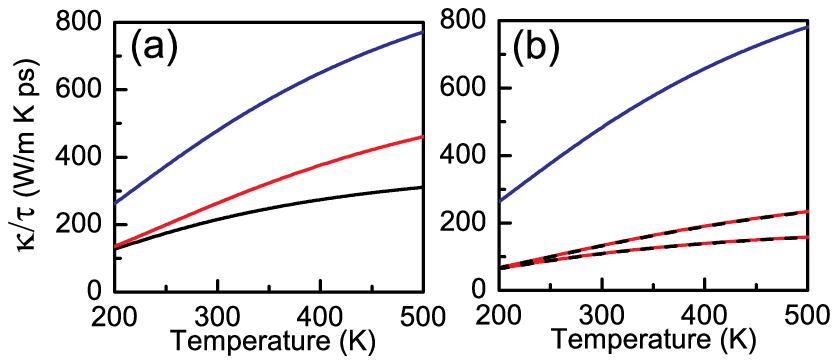}
\label{fig:kappatau}
\end{figure}
\vspace{1in} { Fig. 3(a,b): Kong \emph{et al.} }
\end{center}

\clearpage
\begin{center}
\begin{figure}
\vspace*{0.5 in}
\includegraphics*[width=5.0in]{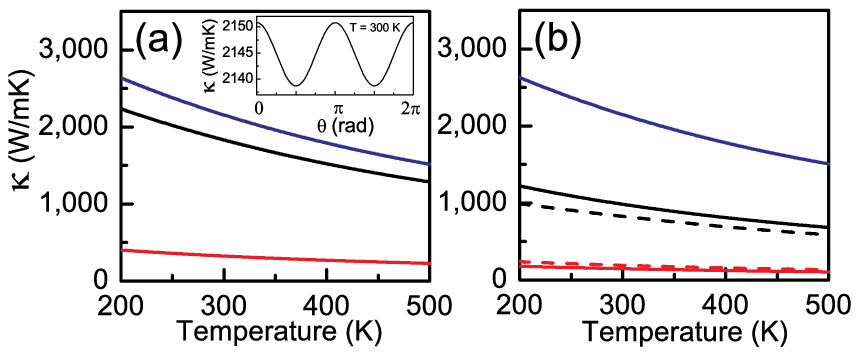}
\label{fig:kappa}
\end{figure}
\vspace{1in} { Fig. 4(a,b): Kong \emph{et al.} }
\end{center}


\begin{references}
\bibitem{thickness1}{We assumed a layer thikness $h = 3.5$~\AA~ for
mono-layer~\cite{thickness2} and $h = 3.5 +3.36 = 6.86$~\AA~ for the bi-layer.}
\end{references}
\end{document}